\documentstyle[12pt]{article}
\newcommand{\ci}[1]{\cite{#1}}
\newcommand{\bi}[1]{\bibitem{#1}}

\begin{document}

\setcounter{page}{0}

\title{ Inclusive Pion Double-Charge-Exchange Above $.5$ GeV}

\author{M.J. Vicente Vacas
\\
{\small Departamento de F\'{\i}sica Te\'{o}rica and IFIC,
      Centro mixto Universidad de Valencia-CSIC,}
\\
           {\small    E-46100, Burjassot, Valencia, Spain}\and
M.Kh. Khankhasayev and S.G. Mashnik
\\
{\small Joint Institute for Nuclear Research, Bogoliubov Laboratory
of Theoretical Physics,}
\\
{\small 141980 Dubna, Moscow Region, Russia}}
\maketitle

\vspace{0.1cm}

\begin{abstract}

A cascade model has been developed to study
pion  induced  multichannel  reactions  (  quasielastic, SCX, DCX,
absorption and $\pi$-production) at  pion energies above $0.5$  GeV.
The  inclusive  pion double-charge-exchange (DCX)
reaction on $^{16}O,^{40}Ca$ and  $^{208}Pb$ nuclei in the  energy
range  from  $0.4$  to  $1.2$ GeV  is  analyzed. Pion energy
spectra and double differential cross section are calculated.  The
pion production is a determinant  feature in the high energy  pion
nucleus  reactions,  and  non  pion  production, DCX signal is sizeable
only at  forward angles and for high energy outgoing pions.
It is  shown that  the contribution  to
inclusive DCX  processes of  the conventional  mechanism, with two
(or  more)  quasielastic  SCX  steps  decreases  very  fast  as  a
function of the energy and reaches very low values at energies above
$.6\,GeV$. This opens the opportunity of having sizeable contributions
of exotic mechanisms that are negligible at the $\Delta$-resonance
energies.

\end{abstract}

\newpage

\section{Introduction}

The pion induced  Double Charge Exchange  (DCX) reaction has  been
extensively  studied  \ci{review1,review2,review3,review4} at
energies below and around the $\Delta$ resonance.  However, many
questions remain  not  sufficiently understood,  like the  angular
dependence of
the  analog  transitions  at  resonance  energy  \ci{o164}, or the
two-peaks structure of the  spectra at forward angles  that appear
in the  inclusive DCX  process in  light nuclei\ci{gram,salcedo,vicente}.
There are many  reasons for these difficulties. One  is
the existence of several mechanisms, some of them important,  that
interfere, and whose evaluation is not simple. Another one is  the
strong  distortion  of  the  pion  waves  at  energies  around the
$\Delta$ resonance.   Finally, for  exclusive reactions,   DCX
processes  are  very  sensitive  to  small  details of the nuclear
structure.

At higher energies  the analysis of  DCX reactions is  much
simpler, mainly, because  the $\pi N$  cross section is  much
smaller,  reducing   the  importance   of   distortion.
Additionally, the contribution  of some of  the mechanisms, like DINT
\ci{dint} and  the  absorption   \ci{absorption}  mechanism, is
expected to decrease rapidly with energy.

Furthermore, the angular dependence of the single charge  exchange
$\pi  N$  reaction  is  strongly  energy  dependent.  This has the
consequence that at  certain energies the  sequential contribution
to the DCX cross section results in deep minima. This result,  found
for the exclusive reaction in \ci{osetdcxhigh} will be shown  here
to be a general   feature of inclusive DCX  for all nuclei.  The
energies, at which  the  sequential  mechanism  is  very  small,  and
distortion is probably negligible,  offer the best ground  for the
investigation  of  some  exotic  mechanisms,  like those involving
meson   exchange   currents   \ci{mikkel}, six-quark    bags
\ci{miller}, and others involving more than two nucleons \ci{jea},
which,   although   small   around   the  $\Delta$
resonance,  are  not  expected  to  decrease  significantly   with
energy.

We know, from  our experience at  low energies, how  strong  is the
dependence of the DCX  cross  section on  the nuclear structure
for exclusive  processes. This dependence is smeared out in  the
inclusive  case,  when  a  sum  over  all possible final states is
done.  Additionally, inclusive DCX offers the advantage of a  much
higher yield than exclusive experiments and does not require  such
a good energy resolution.

At  present  there  are  no  data  on inclusive DCX in the energy
range above $.5\,GeV$, although some measurements are in  progress
now \ci{Kuli} at ITEP  (Moscow). These experiments have stimulated
the present work. There are, however, some data around .5 GeV which
will be used to test our model.

The  calculation  presented   here  is  a  Monte   Carlo
simulation  of  the  reaction.  This  is  most  appropriate  for a
multichannel situation (i.e.  absorption, quasielastic, SCX,  DCX,
$\pi$-production)  that   renders  a   full  quantum   calculation
infeasible.

Furthermore, similar  calculations \ci{salcedo,vicente}  describe
quite  well  all  $\pi$-nucleus   inclusive  reactions  around the
$\Delta$-resonance:  absorption,  quasielastic  scattering, single
charge  exchange,  and  double  charge  exchange.  At higher
energies  the  wavelength  of  the  pion  is shorter and therefore
quantum interference effects should be less important.

This  paper is  organized as  follows:  Section 2  presents
the basic elements  of the  variant of the   cascade model
considered here (details  are  given  in  Appendix  A). In Section 3
the results of calculations of pion spectra (energy and angular
distributions) for the inclusive DCX are  presented, and in Section 4
the  obtained  results are discussed.

\section{Model}

\subsection{Basic considerations}

A pion  travelling inside  a nucleus  can be  absorbed, can change
direction, energy, charge, or  even produce more pions.  The basic
inputs  for  our  simulation  will  be  the probabilities per unit
length for each of these channels  to happen.  How  these
probabilities are obtained is presented below,
in sections 2.2 -  2.4. Details
on the simulation will be presented in Appendix A.

\subsection{$\pi N \longrightarrow \pi N$}

The probability  per unit  length of  quasielastic scattering  (or
single charge exchange) is given by

\begin{equation}
P_{N(\pi^{\lambda} ,\pi^{\lambda '})N'} =
\sigma_{N(\pi^{\lambda} ,\pi^{\lambda '})N'}\;\rho_N
\end{equation}

\noindent
where  $N$   stands    for  neutron   or    proton,  $\rho_N$   is
the density  of        nucleons       of         type         $N$,
and  $\sigma_{N(\pi^{\lambda},  \pi^{\lambda   '})  N'}$        is
the   elementary      cross     section    for   the      reaction
${\pi^{\lambda} +  N \rightarrow   \pi^{\lambda   '}+N'}$ obtained
from Arndt's   phase shifts   \ci{Ar}.     The       density    of
protons  is taken  from   experiment,   and    the        density
of  neutrons   is   taken proportional   to   the     density   of
protons  in   all  results presented in this work.

When according   with    the  probabilities   of  eq.  (1),   a
quasielastic  scattering  took place,  we  executed the following
algorithm. First, we  chose randomly a  nucleon, of
the type $N$, from  the  fermi   sea, then we boosted  the $\pi$ and
$N$ to their center of mass system.   Finally,  we   selected  the
scattering angle   (and   therefore   energy)   of the outgoing
particles  using  again    the   experimental    cross    sections
\ci{Ar},  and boosted the  momenta to  the lab.   system.   When  the
momentum of  the outgoing nucleon in the lab. system is  below  the
fermi  level,  we   consider  the  event  to  be Pauli-blocked and
therefore keep the pion  initial charge and momentum unchanged.

\subsection{ Pion absorption}

Even  if  pion  absorption  is  a  relatively small effect at high
energies (which one could suggest  from the rapid decrease of  the
pion-deuteron absorption  cross section  at the  energy range from
$0.3\,-\,1.0\,GeV$  \ci{akemoto,borkowski}),  there  is  a   large
number of pions at lower energies which are generated both by  the
quasielastic rescatterings and the pion production on the  nuclear
nucleons.   The proportion  of these  pions that  eventually comes
out of the nucleus  is essentially  determined by  the absorption
strength.

Although  pion  absorption  has  been  extensively studied at
energies below  0.3 GeV, little theoretical  work has been done
about pion absorption  by complex nuclei  at high  energies,
and very little experimental information is available.
In ref.  \ci{osetscxhigh} the
effect of   pion absorption  on the  pion-nucleus elastic  and the
SCX scattering  has been  studied in  the energy  range of  250 to
$650\,MeV$.  The  absorptive  part  of  the  pion-nucleus  optical
potential  was  calculated  within  the  framework  of a many-body
field  theoretical  approach.   The  model  contains both two- and
three-nucleon  absorption  mechanisms  and  it  has  been shown to
agree quite well to the  more complex microscopical model of  ref.
\ci{futami} in the $\Delta$ resonance region.  Results  show
a quite  weak absorption  at high  energies, as  was  expected.
Another interesting  result  presented  in  ref.\ci{osetscxhigh} is
that,  whereas  in  the  resonance  region   three-body-absorption
becomes comparable  with the  two-nucleon mechanism,  as the pion
energy  increases  the  effects  of three-body absorption decrease
again, and  the two-body  mechanism becomes  dominant, as  is the
case at low energy.

The probability per unit length of a pion to be absorbed is
expressed in terms the imaginary part of the pion self-energy
\footnote{The pion selfenergy $\Pi$ is  related to the optical
potential  as $V\,=\,{\Pi}/2k^0$} by the equation
$P_{abs}=-Im\Pi_{abs}(k)/k$.
The  imaginary    part  of    the  pion    self-energy,
related to  two-nucleon   pion  absorption,
which  has  been  calculated in \ci{osetscxhigh} is of the form

\begin{equation}
Im\Pi_{abs}^{(2)}(k)\, = \,-D_2 {{s\,\bar\sigma}\over{ q}}
\,{ \rho}^{2}
\end{equation}
Here, $D_2\,=\,0.0116\,fm^5\,mb^{-1}$,  $k$ is  the pion  momentum
in the  lab system,  $s$ is  the square  of the  c.m.  energy of a
pion   of   momentum   $k$   and   a   nucleon   at  rest,  ${\bar
\sigma}\,=\,({\sigma}_{3/2}   +    {\sigma}_{1/2})/3$    is    the
spin-isospin averaged  unpolarized $\pi N$ cross section, and  the
momentum of  a virtual  pion $q$  that appears  in the  model of ref.
\ci{osetscxhigh} is determined as

$$
q=\{[\frac{(k^{0}+2m)^2\,-\,\vec{k}^2}{2(k^{0}+2m)}]^2
-\,m^2\}^{1/2}
\nonumber
$$
which  in  the  nonrelativistic limit  actually used in \ci{osetscxhigh}
goes to

$$ q=[m(k^{0}-\vec{k}^{2}/{2m})]^{1/2}
\nonumber
$$
where $m$ is the mass of a nucleon.

The   three-nucleon   pion   absorption    has   been   calculated
\ci{osetscxhigh} in  a similar  fashion, and  the contribution  to
the selfenergy is given by

\begin{equation}
Im\Pi_{abs}^{(3)}(k)\, = \,-D_3 {{s\,\bar\sigma}\over{ q'}}
\,{{s'\,\bar\sigma'}\over{ q'}}
\,{ \rho}^{3}
\end{equation}

\noindent
with

$$
q'=[{{m}\over{2}}(k^{0}-\vec{k}^{2}/{2m})]^{1/2}
\nonumber
$$
and $s'$,$\bar\rho'$ have the same meaning as $s$ and  $\bar\rho$,
but are evaluated  at a kinetic  energy of the  pion equal to  two
thirds of the real one ($T_\pi'={{2}\over{3}}T_\pi$), and
$D_3=1.15\cdot 10^{-7}$ fm$^8$mb$^{-2}$MeV$^{-1}$.

The pion  selfenergy pieces  of eqs.  (2) and  (3) can  readily be
translated  to a probability per unit length by the relation
$P_{abs}=-Im\,\Pi_{abs}(k)/k.$

In  \ci{arima}  (see,  also  Erratum  in  \ci{arima1}),  the  pion
absorption  effect  on  the  pion-nucleus scattering at $800\,MeV$
has been estimated  using the quasi-deuteron  (qd) model.   Taking
into account that at energies well above the resonance region  the
two-body  absorption  becomes  dominant  \ci{osetscxhigh},  it  is
interesting to compare  the outlined above  model (Eq.(2)) with  the
quasi-deuteron model.

The imaginary part of the pion selfenergy $Im\Pi_{abs}(k)$  within
the framework of the quasi-deuteron model is given by

\begin{equation}
Im\Pi_{abs}^{(qd)}(k)\,=
-\,\,8{\pi}{\Gamma}\,
ImB_{\pi 2N}(E){\rho}_{n}{\rho}_{p}
\end{equation}
where ${\rho}_{p}$ and ${\rho}_{n}$  are the densities of  protons
and neutrons.
 The kinematical factor $\Gamma $ is determined as
$$
\Gamma \,=\,\gamma\,\frac{{\cal M}_{\pi A}}{{\mu}_{\pi d}}
\nonumber
$$
where,  ${\cal   M}_{\pi  A}$   and   ${\mu}_{\pi   d}$  are   the
$\pi$-nucleus and  $\pi$-deuteron reduced  masses correspondingly,
and $\gamma  $ is  the relativistic  transformation factor  of the
$\pi -2N$  scattering amplitude  from the   $\pi $-nucleus  to the
$\pi -2N$ c.m.s.

$$
\gamma =[{\omega}_{\pi}({\kappa}){\omega}_{\pi}({\kappa}' )
E_{2N}({\kappa})E_{2N}({\kappa}')/
{\omega}_{\pi}(q){\omega}_{\pi}(q' )
E_{2N}(P)E_{2N}(P' )]^{1/2}
\nonumber
$$
Here, $\vec{q}$  and $\vec{q}'$  are the  pion momenta  before and
after the collision in  the $\pi $-nucleus c.m.s.;  $\vec{\kappa}$
and $\vec{\kappa}'  $ are  the pion  momenta before  and after the
collision in the  $\pi -2N$ c.m.s.,  and $\vec{P}$ and  $\vec{P}'$
are the  total  momenta of the $2N$  subsystem  before  and after
the collision in the $\pi $-nucleus c.m.s. We  use here  the "frozen"
approximation which  means that $\vec{P}\,=\,-2\vec{q}/A$  and
$\vec{P}'\,=\,\vec{P}\,-\,{\vec{\Delta}}$,  where
${\vec{\Delta}}\,=\,\vec{q}'\, -\,\vec{q}$ is the momentum transfer.

The imaginary part of the absorption parameter $B$ is given by

$$
ImB=(1/{4\pi})W(T_{\pi})/(2{\rho}_d(0))
\nonumber
$$
where  ${\rho}_d(0)$  is the deuteron  density at $r=0$.  So as in
\ci{arima}  this  quantity  is  calculated  using  the  square-well
potential model of  a deuteron.   The parameter $W$  is related to
the pion-deuteron total cross section as
$
W\equiv q\*\sigma ({\pi^{+}d\rightarrow pp})
\nonumber$.
 The total pion-deuteron  absorption cross section   at the  energy
range  from   $0.3\,GeV$  to   $1\,GeV$  can  be  obtained  from   the
differential  cross  section  data  of \ci{akemoto,borkowski}. The
energy dependence of $W(T)$ can be approximated as

$$
W(T_{\pi})={\alpha}_1/T_{\pi}+{\alpha}_2/T^{2}_{\pi}+{\alpha}_3/T^{3}_{\pi}
\nonumber
$$
where  $T_{\pi}$  is  the  pion  kinetic  energy in the laboratory
system (measured in fm), and the parameters $\alpha$ are

$$
{\alpha}_1\,=\,0.171,\,\, {\alpha}_2=-0.612fm^{-1},\,\,{\alpha}_3
=\,1.780fm^{-2}
\nonumber
$$

\subsection{ Pion production}

Pion production is a determinant feature  in the high energy  pion
nucleus reactions. The  inelastic channels have a  cross
section comparable, or  even larger than  the elastic channels  at
energies above 0.6  GeV.  Although  two (or more)  pion production
channels  are  possible  at the  energies considered, the
inelastic cross section  is clearly dominated  by the single  pion
production \cite{lbl}.  In this work the  multipion production
channels have been ignored  \footnote{ Note that this omission will  not
affect practically  the higher  energy part  of the  pion spectra,
because for these channels some energy,  at least two pion masses,
has  necessarily  been spent.}.

Whereas at low energies a considerable amount of data is available
for most isospin channels, including differential cross  sections,
our knowledge is more fragmentary in the energy regime addressed
here. Data has been taken  from compilation  \ci{lbl} and
\ci{Otro,Manley},  to   obtain  parametrizations   of  the    $\pi
N\rightarrow  \pi\pi  N$  total  cross  sections.  Then,  for each
channel, the probability per unit length is given by the equation

\begin{equation}
P_{N(\pi,2\pi)N'} =
\sigma_{N(\pi,2\pi)N'}\;\rho_N
\end{equation}
When, according to this probability, a pion production event of  a
certain  isospin  channel  has  taken  place,  we  proceed  in  the
following way.  First, a nucleon of the type $N$ is  randomly chosen
from  the   fermi sea.  Then the scattering angles
and energies of the outgoing particles are selected, using the 3-body
phase  space  distribution.   When  the  momentum  of the outgoing
nucleon in the lab system is below the fermi sea level,
the event is considered to be Pauli-blocked and therefore the
pion  initial charge and momentum are unchanged, and no
any new pions are  produced.

There are five  independent pion production   channels induced  by
charged pions, namely,
$$(1)\quad\pi^++p\rightarrow\pi^+\pi^+n$$
$$(2)\quad\pi^++p\rightarrow\pi^+\pi^0p$$
$$(3)\quad\pi^++n\rightarrow\pi^+\pi^0n$$
$$(4)\quad\pi^++n\rightarrow\pi^0\pi^0p$$
$$(5)\quad\pi^++n\rightarrow\pi^+\pi^-p$$
The cross sections of (1) and (2) are parametrized as

$$\sigma_{(1)}=\sigma_{in}\,0.2\,[1-0.05\,(T_{\pi}\,-
\,T_{\pi th})],$$

$$\sigma_{(2)}=\sigma_{in}\,0.8\,[1-0.05\,(T_{\pi}\,
-\,T_{\pi th})],$$
where  $\sigma_{in}$  is  the  inelastic  $\pi^++p$  cross section
obtained  from  Arndt's  phase  shifts,  $T_{\pi}$  is the kinetic
energy of a pion in units of $fm^{-1}$ and $T_{\pi th}$ is  the
kinetic  energy  of  the  pion  production  threshold  in the same
units.

In a similar fashion the other three channels have been parametrized.
The formulas are

$$\sigma_{(3)}\,=\,(\sigma_{in} -\sigma_{(4)}-\sigma_{(5)})
\,[1-0.075(T_{\pi}-T_{\pi th})],$$

$$\sigma_{(4)}\,=\,\sigma_{in}\,[0.30-5.069\times 10^{-2}
T_{\pi}+5.004\times 10^{-3}T_{\pi}^2]$$

$$\sigma_{(5)}=\,\sigma_{in}\,[-2.78\times 10^{-2}+0.315 T_{\pi}
-4.154\times 10^{-2}T_{\pi}^2]$$

where $\sigma_{in}$ is now the inelastic $\pi^++n$ cross  section,
and $T_{\pi}$,\,$T_{\pi th}$  have the  same meaning  as above. In
all these cases the reproduction of data is quite satisfactory.

Apart  from  the  five  additional  channels induced by a $\pi^-$
that  will  be  obtained  by  charge  symmetry,  there  are  three
channels induced by a $\pi^0$:

$$(a)\quad \pi^0+N\rightarrow\pi^0\pi^0 N$$
$$(b)\quad \pi^0+N\rightarrow\pi^+\pi^- N$$
$$(c)\quad \pi^0+N\rightarrow\pi^0\pi^c N'$$
where $N$ is a nucleon, $\pi^c$ is a charged pion, and the
charge of the nucleon $N'$ is determined by  the charge balance
in the channel (c).
 We use $\sigma_{(a)}=\sigma_{(b)}=\sigma_{(c)}=\sigma_{in}/3,$
where $\sigma_{in}$ is the $\pi^0N$ inelastic cross section.
These  channels are  less  important,  because the
experiment will  begin  with  a  charged  pion. Therefore the
neutral pions  are  secondary  pions. That  means that
there are fewer of them, and  also that in the average they have  less
energy  and  will not affect  much the  higher  energy part of the
spectra of interest here.

\section{ Results}
There are some data of pion-nucleus scattering at energies  around
0.5 GeV.  We compare  the results of our  program
with these  data, because  data at  higher energies  are yet  very
preliminary.  We should remark however that we expect our  results
to  be  in  better  agreement  with experiment at higher energies,
where the use  of a semiclassical  approximation like this  one is
more justifiable, and where the pion interaction with nucleons  is
weaker.

In  Fig.1  we  show  the results of the present model (solid line)
for the quasielastic
$\pi^+$  scattering  in  $^{12}C$,  compared with the experimental
data from ref.   \cite{Zumbro}.

The  quasielastic    peak   is
well reproduced,     in both    size    and    width,     at   all
angles.   The   size,  being  absorption   of   very    little
importance  at  this  energy in  our model,  is governed   by  the
elementary    $\pi N$  cross      sections  and by   the     fermi
motion   of    the   nucleons.
The  calculation   we  present   has  been  done assuming a  fermi
momentum of 250  MeV all over   the nucleus.   In fact, the    use
of   a   local   fermi   momentum,   obtained    from the    local
nucleon  density,   results    in    a    worse  agreement
with     the     data,   producing     narrower     peaks      and
overestimating  clearly the cross section at forward angles.

Our  results underestimate  the  cross  section  at pion energies
below  the quasielastic  peak. This  seems to  be a  fact common
to  other  cascade  codes (see, discussion in ref. \cite{Zumbro}
and Sect. 4)

Of course  we cannot reproduce    the elastic    peak, important
at low    angles, given    that we  do  not  have    collisions of
the pions  with the nucleus as  a whole.  Only pure   quasielastic
scattering  is    our aim  here.

In Fig. 2 we show   the results of the present model  (solid line)
for single  charge  exchange   scattering at 500 MeV.   This  is a
very  important  channel    for   us,    because  double    charge
exchange  requires    two  single   charge  exchange  scatterings.
Again,  the    quasielastic     peak size  and  width  are in    a
quite  good   agreement with   data. This   should   be  expected,
as       this    peak        is   dominated     by   a      single
pion-nucleon   collision,    as    it   was    the      case   for
quasielastic scattering.     The only  additional information   is
that the  ratio between    different scattering channels is    the
same as in     the  elementary    $\pi    N$ collisions.      Note
also that   below  the  peak, one can observe  the   same behavior
found in      quasielastic scattering.   There   are   some  pions
missing   in   that   region.   In   this   figure,   we      show
separately    the   pions     coming  from     $\pi$   production,
which are   not   enough   to     agree with the data from
\cite {Pet}.  Even  the  total  suppression  of pion   absorption  would
not  be   enough     to  improve  significantly the agreement with
data.

In Fig.2 we also compare the data with the cascade-exiton model (CEM).
This model of nuclear reactions \cite{cem} was proposed initially
to describe nucleon induced reactions at bombarding energies below
or at $\sim 100$ MeV and was later developed for a larger interval
of bombarding energies and for the analysis of pion-nucleus reactions
(see, e.g. \ci{Mashnik} and references therein.)
The  results  of  CEM, shown by the short-dashed curves,
agree with  the results  of the  present model.  The  CEM
calculations also underestimate  the low energy  part of the pion
spectra.

Finally, before concentrating in the DCX channel, we will  present
some results  for pion-nucleus  reactions at  higher energies. Our
purpose is to  identify the main  features of these  processes. In
Figs.3-5  we   analyze   the  reaction
$\pi^+  +   ^{40}Ca \rightarrow  \pi  +  X$,  and  Figs.6,7  deal
with the same
reaction in lead.   In all figures we   split the total  cross
section into two pieces: a  quasifree piece, given by those  pions
that  come  out  of  the  nucleus  after  having  only   quasifree
scatterings,  and  a  pion  production  piece,  given by the pions
coming  from  events  in  which  at  least  a pion production took
place.   Note  that  in  this  latter case, quasifree scatterings,
prior  or  subsequent  to  the  pion production itself, could have
occurred.

Fig.3 shows the energy  spectra of the outgoing pions.   Although
the energy  spectra begin  at 0  MeV, we  should remember that the
results  at  very  low  energies  are  not meaningful, because the
cascade method is   not appropriate there.  In ref.  \cite{salcedo}
it  has  been  shown  that  quantum  calculations  begin to differ
appreciably from the present  kind of approach, and  for inclusive
processes at energies around 100 MeV.

Let us begin  discussing  the  quasielastic channel.   In it,   we
can   separate  a   region  of   high  energies,  where only pions
coming from  one (or  several)   quasifree scatterings contribute,
and  a   second  region   dominated  by   pions  coming from $\pi$
production events. Note   the little dip   around 150 MeV  in  the
"quasifree" part, and  also the change  of  curvature  in the same
region of  the   "production" part.  Both are  due to   the strong
absorption  in the  $\Delta$ resonance  region. At energies  above
300  MeV  absorption  effects  are small for both the two models
discussed before.

The  situation  is  similar  for  the  SCX  channel,  although the
quasifree peak is smaller.   As expected, because DCX requires  at
least two scatterings, the quasifree  peak is much smaller in  its
case.

The importance  of pion  production channels  contribution to  DCX
has already been shown at lower beam energies \cite{vicente}.   At
600 MeV and above, DCX  is totally dominated by $\pi$  production.
Of  course,  this  is  true  except  for the small region of phase
space where $\pi$ production is forbidden.  Therefore, if we  want
to learn  something about  alternative DCX  mechanisms, we  should
concentrate our attention in  this region.  Otherwise,  any signal
will  be  blurred  by  the  large  number  of  pions  coming  from
production channels.  And  these production channels are  not well
known,   neither in  size, nor  in the  angle-energy dependence of
its cross sections.

The  angular  behavior  of  the  reaction  is shown in Fig. 4. All
channels  are  forward  peaked,  essentially the "quasifree" part.
Note that one of the  uncertainties in our model is,  as mentioned
above,  the   angle-energy  structure   of  the    $N(\pi,2\pi)N'$
amplitudes, that has been included as being proportional to  phase
space. Thus, the  "production" part reflects  mostly the boost  of
the isotropic process from the CM to the lab system.

We have  selected the  DCX channel  to show  a double differential
cross  section  in  Fig.  5.  Given  the angular behavior observed
previously  it  is  only  logical  to  find that the best place to
isolate a  clean, non-production,  DCX signal   occurs at  forward
angles and for high energy pions. The rest of the spectra are totally
dominated by the $\pi$-production channels.
 Observe  again  the  dip  produced  by  the absorption at
energies around the $\Delta$ resonance.

Fig. 6 shows the pion  spectra for the reaction $\pi^+  + ^{208}Pb
\rightarrow \pi  + X$  at 1200  MeV. Many  of the  main
features found in  Calcium at 600  MeV are also  relevant for this
case.  Let us mention that there is an even stronger dominance  of
the production  channels, except  for the  quasielastic channel at
the very high  energy region. The  reason is the  smallness of the
$\pi  N$   charge  exchange   cross  section   at  this   energy.
Obviously, this implies a small "quasifree" SCX cross section  and
an even smaller "quasifree" DCX cross section.

The angular distribution, not  presented here, is similar  to that
of Calcium, although  a bit more  forward peaked,   essentially in
the  quasielastic   channel.  Fig. 7  shows   the  DCX    double
differential cross section for the reaction $\pi^+  + ^{208}Pb
\rightarrow \pi  + X$  at 1200  MeV
at the same angles presented before for Calcium.
Quasifree DCX is hardly visible with the scale used in  the figure,
as it could be expected from Fig. 6.

Finally, we have selected as observable the integrated DCX  cross section,
putting as a cut  that the energy of the final pion is,  at most,
150 MeV  below the  beam energy.  This eliminates  practically all
cases  in  which  there  is  a  pion  production.  The results, in
Calcium,  and  as  a  function  of  the  beam  kinetic energy, are
presented in Fig. 8. Very similar results are obtained putting  an
additional  cut  in  angles,  given  that most pions fulfilling the
previous  energy  condition  go  forward.   For comparison we also
present the  quasielastic and  SCX case.  Note that  whereas the
quasielastic  channel changes by around a factor two
(from 500 to 1300 MeV), SCX  loses  one  order of magnitude and
DCX decreases almost 3 orders of magnitude.

We do  not consider  this as a  prediction of an extremely  low DCX
cross section at high energies. We do not expect, nor claim  that.
First, because  of a  "technical" reason,  DCX is  so tiny because
quasifree SCX is quite  small. A change of  the SCX $\pi N$  cross
section by  a 2\%  of a  typical scale,  say the  elastic $\pi  N$
cross section,  would mean  a factor  4 for  our result.  In other
words, the error bars are only statistical and do not include  the
errors coming from the elementary cross sections used as input.  A
second, more important  point is that  the curve corresponds  only
to  the  ingredients  of  the  code,  namely,  to  two consecutive
quasifree single charge exchange $\pi N$ collisions. Let us  them
state  our  result  in  a  meaningful  way:   The  contribution to
inclusive DCX  processes of  the conventional  mechanism, with two
(or  more)  quasielastic  SCX  steps  decreases  very  fast  as  a
function of the energy and reaches very low values, compared  with
the  quasielastic channel, at energies above 600 MeV.

\section{ Discussion}

In the present paper a cascade model has been developed
to study pion induced multichannel reactions (quasielastic, SCX, DCX,
absorption and $\pi$-production) at pion energies above $.5$ GeV.

The   model has  been checked  by comparing  with the experimental
data  \cite{Zumbro,Pet}  for  quasielastic  pion   scattering  and
single  charge  exchange  $\pi^-$  scattering  in  $^{12}C$ at $0.5
GeV$.  We also compared the results of the present model with  the
CEM \cite{cem,Mashnik} calculations.

It  has   been  shown   that  for   quasielastic  scattering   the
quasielastic peak is well reproduced,  in both size and width,  at
all  angles,  but  our  results  clearly  underestimate  the cross
section at pion energies below the quasielastic peak.  This  seems
to be a fact  common to other cascade  codes as it is  remarked in
ref \cite{Zumbro}, indicating possibly, some interesting piece  of
physics missing in our description of the pion-nucleus  reactions
\ci{wise}. The  CEM  calculations  also  underestimate  strongly
the   cross section at this energy range.

For the single charge exchange  scattering we  also observe   some
missing of pions below the quasielastic peak.

One  could  ascribe  the  missing  cross sections to several causes
that should  certainly be  investigated further.  For instance, at
lower energies, a sizeable  enhancement of the ($\pi,2\pi)$  cross
sections,  when  comparing  to  quasifree  calculations,  had been
predicted  in  ref.  \cite{vi2pi}  and  was  later  found  in ref.
\cite{grion}.  The  effect  was  related  to  the  change  of  the
dispersion  relation  of  the  pions  in  the  medium. Speaking in
simple  terms,   the    pions  are   attracted  by   the   medium.
Unfortunately  we  do  not  know  so  well  the  pion  propagation
properties  at  higher   energies,  and  the   results  of    ref.
\cite{vi2pi} cannot be easily extrapolated. In ref.  \cite{Zumbro}
Zumbro et al. suggest the  formation of a narrow $\sigma$   meson,
with little  interaction   with the  medium, that  would leave the
nucleus prior  to its  decay into  two pions.   One should mention
that  the  obvious  candidate,  a  weaker  pion  absorption, it is
difficult  to  reconcile  with  the  agreement  obtained  in   the
quasielastic peak, that  for different angles  is situated at  the
same  energy,  and  also  with  the  pion  absorption  data in the
resonance energy region.

At  $.6\,GeV$  and  above,  DCX  is  almost totally  dominated  by
$\pi$ production.  To learn something about exotic DCX  mechanisms,
we should concentrate our attention  in the region of phase  space
where $\pi$ production is  forbidden.  Otherwise, any  signal will
be blurred  by the  large number  of pions  coming from production
channels.  Furthermore,  these  production  channels  are  not well known,
neither in size, nor in  the angle-energy dependence of its  cross
sections.

The calculations for the inclusive DCX at energy $1.2\,GeV$  show that
many  of the  main
features found  at 600  MeV are also  relevant at these higher energies.
 There is an even stronger dominance  of
the production  channels, except  for the  quasielastic channel at
the very high  energy region. The  reason is the  smallness of the
$\pi  N$   charge  exchange   cross  section at these energies.

There are  some interesting  results in  the literature concerning
exclusive  DCX  processes  at  high  energies.  In  particular the
reactions $^{18}O(\pi^+,\pi^-)^{18}Ne$  and also
$^{14}C(\pi^+,\pi^-)^{14}O$   have    been   studied    in    ref.
\cite{osetdcxhigh} at  energies up  to 1400  MeV. Their  resulting
cross sections present two deep minima at energies around 700  and
1300 MeV.   That result  does not  depend on  nuclear structure or
the  specific  nuclei  chosen.  It  simply  reflects  the   energy
dependence of the $\pi N$ SCX  amplitude.  Thus one expect to  get
a  similar  result  for  the  inclusive DCX process and, possibly,
gaining in yield and  requiring a less precise  energy measurement
of the final pion, because there is no need to separate clearly  a
given final state of the target nucleus. To study this point,
 we have selected  as observable the integrated  DCX cross
section (Fig. 8), putting  as a cut that  the energy of the  final
pion is, at most, 150 MeV below the beam energy.  This  eliminates
practically all cases in which  there is a pion production.   Very
similar results are obtained putting an additional cut in  angles,
given that most pions  fulfilling the previous energy  condition go
forward. Whereas the  quasielastic  channel    loses   less   than
one  order of magnitude (from $.5\,GeV$  to $1.3\,GeV$), SCX
changes  by a factor  1/40, and DCX decreases 3 orders of magnitude.

It should be noted that, within the present approach, the  contribution to
inclusive DCX  processes is generated by the conventional  mechanism,
with two (or  more)  quasielastic  SCX  steps. The results presented
in Fig.8 show that the sequential mechanism  decreases  very  fast  as  a
function of the energy and reaches very low values, compared  with
the  quasielastic channel, at energies above 600 MeV. In this situation,
it is very important to have a possibility to compare these results
with the experimental data and to consider other mechanisms of the
DCX (like MEC) which might do not decrease so fast at high energies.

\section*{Acknowledgments}

This work was partially supported by CICYT, contract AEN  93-1205,
and, in  part by  the Russian   Science Foundation  (Project   no.
93/94-02-3833).  M.  Kh. would like  to  express  his thanks for  the
hospitality  extended  by  University  of  Valencia.  We  are also
grateful to E.Oset for discussions and help.

\newpage

\appendix
\section{ Details on the simulation}

We generate pions, of a  given momentum and charge, which  travel
in the  $z$ direction,  with a  random impact  parameter $\vec b$,
obeying $\vert\vec  b \vert\leq  R$, where  $R$ is  an upper bound
for the nuclear  radius. We choose  $R$ such that  $\rho(R)\approx
10^{-3}\rho_0$, with $\rho_0$  the normal nuclear  matter density.
At the  beginning, the  pions are  placed at  the point  $(\vec b,
z_{in})$, with $z_{in}=-\sqrt{R^2-\vert  b\vert^2}$, and then,  we
proceed to  move them  along the  $z$ direction,  in small  steps,
until either the pions get out of the nucleus or interact.

Let  us  assume  that  $P(q,r,\lambda)$  is  the  probability   of
interaction  per  unit  length,  at  the  point  $r$, of a pion of
momentum $\vec  q$ and  charge $\lambda$.   We choose  an interval
$\delta l$, such that  $P(q,r,\lambda)\delta l$ is small  compared
to unity. Then we generate  a random number $x\in[0,1[$.   We have
two possibilities:

{\sl (a)} $x>P\,\delta l$. In  this case there is no  interaction,
and the pion travels a distance $\delta l$ along the direction  of
its momentum $\vec q$.

{\sl  (b)}  $x<P\,\delta  l$.  In  this case there is interaction.
According  to  its  respective  weights,  we decide whether it has
been  absorption,  quasielastic  scattering,  charge  exchange, or
pion production.

When it  has been  quasielastic, or  charge exchange,  we use  the
procedure  defined  in  section  2.2  to  find the new energy, and
direction of the pion, and continue to propagate it along its  new
direction, checking at every step if new interactions take place.

When it has been a pion  production case, we choose the   channel,
according  with  the  respective  weights  given  by  their  cross
sections, and  select the  energy and  direction of  the two final
pions  using  the  algorithm  described  in  section 2.4.  Then we
store the data of one of the pions and keep moving the other one.

In the case that  after moving the step  $\delta l$ the pion  gets
out  of  the  nucleus,  and  when  the  pion is absorbed, we check
whether there are some other pions left inside the nucleus  (these
pions would have been produced previously at some step). If  there
are  some  other  pions,  we  select  one  of  them  and  begin to
propagate it from its current position.

When  there  are  no  pions  left,  we  store the data -energy and
angles- of  all pions  that got  out of  the nucleus,  if any, and
begin again the full procedure by generating a new initial pion.

Note   that   between   interactions   pions   follow  a  straight
trajectory.  Thus, even some classical effects like the change  in
direction  due  to  the  real  part  of  the potential ,strong and
coulombian, are neglected.

\subsection{Integrated cross sections}

Let $N$ be the  total number of incident  pions, and let $N_c$  be
the total  number of  events of  a given  channel, i.e.  number of
cases in which there  is a $\pi^-$ in  the final state, then,  the
integrated cross section for that channel is given by

$$
\sigma_c =\pi R^2 {{N_c}\over{N}}
$$

\subsection{Differential cross sections}

As  an  example,  we  will  show  the  way  in which angular cross
sections  are   calculated.  Energy   distributions,  or    double
differential  cross  sections  are  obtained  in a similar way. To
calculate  differential  angular  cross  sections  we  divide  the
cosine of the polar angle in $N_\mu$ equal angular intervals.   If
$\mu$ is the cosine  of the polar angle  with which a pion  leaves
the nucleus, we associate to it the discrete value of the  angular
bin in which it falls. Thus

$$k=1 + [{{\mu+1}\over{\delta\mu}}]$$
where, [.] means that we take only the integer part, and
$$\delta\mu=2/N_{\mu}.$$
If we get a total number of $n_k$ pions in the $k$ bin, we have

$$
{{d\sigma_{k}}\over{d\Omega}}=  \pi R^2 ({{1}\over{2\pi\delta\mu}})
{{n_k}\over{N}}
$$

\newpage

\section{Figure Captions}
\bigskip
\parindent 0cm

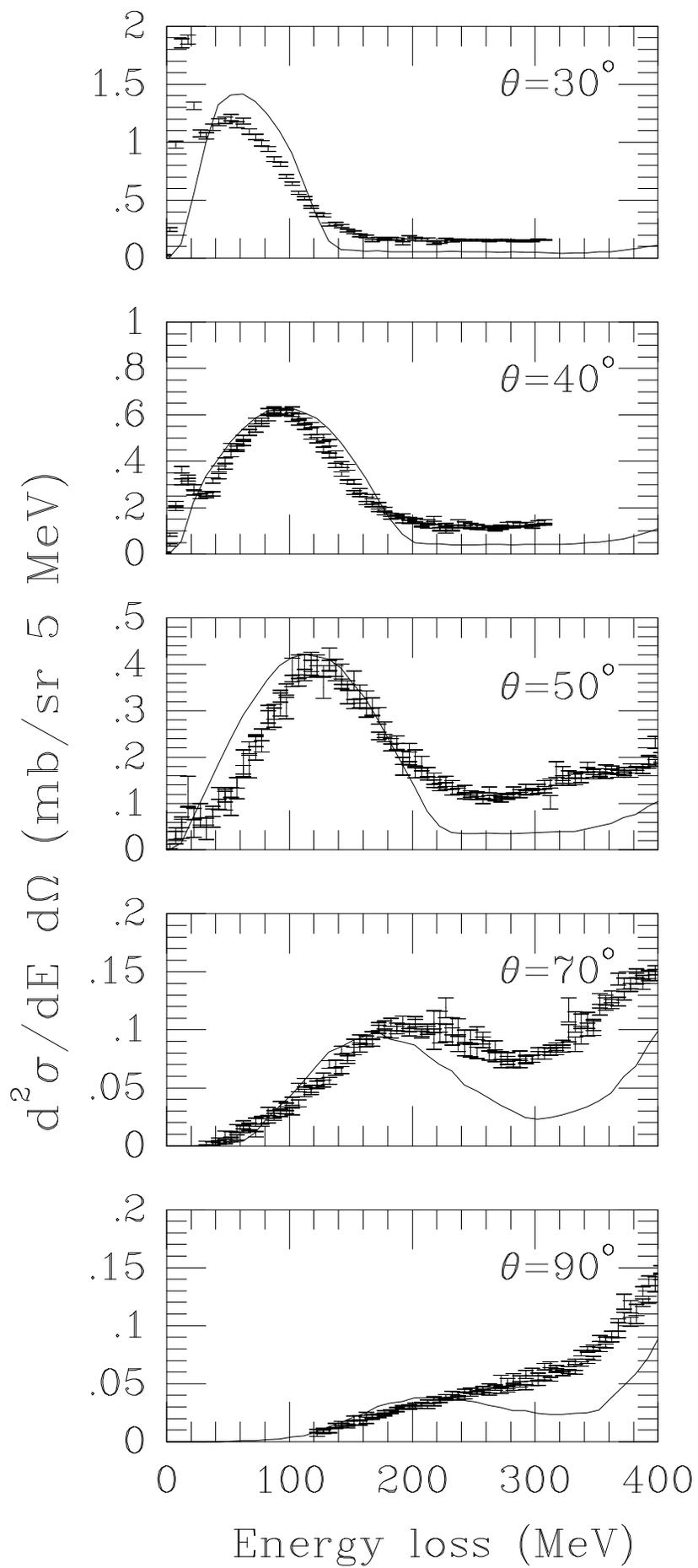
{\bf Fig. 1.}  Double differential cross section for the
$^{12}C(\pi^+,\pi^+)X$ reaction at T$_{\pi}$=500 MeV.
Continuous line: full model.The data, taken from ref. \cite{Zumbro},
also include the $^{12}C(\pi^-,\pi^-)X$ reaction.

\medskip
{\bf Fig. 2.}  Double differential cross section for the
$^{12}C(\pi^-,\pi^0)X$ reaction at T$_{\pi}$=500 MeV.
Continuous line: full model, long-dashed line: $\pi$-production
channels, short-dashed line CEM model \cite{Mashnik}.
Data from ref. \cite{Pet}.

\medskip
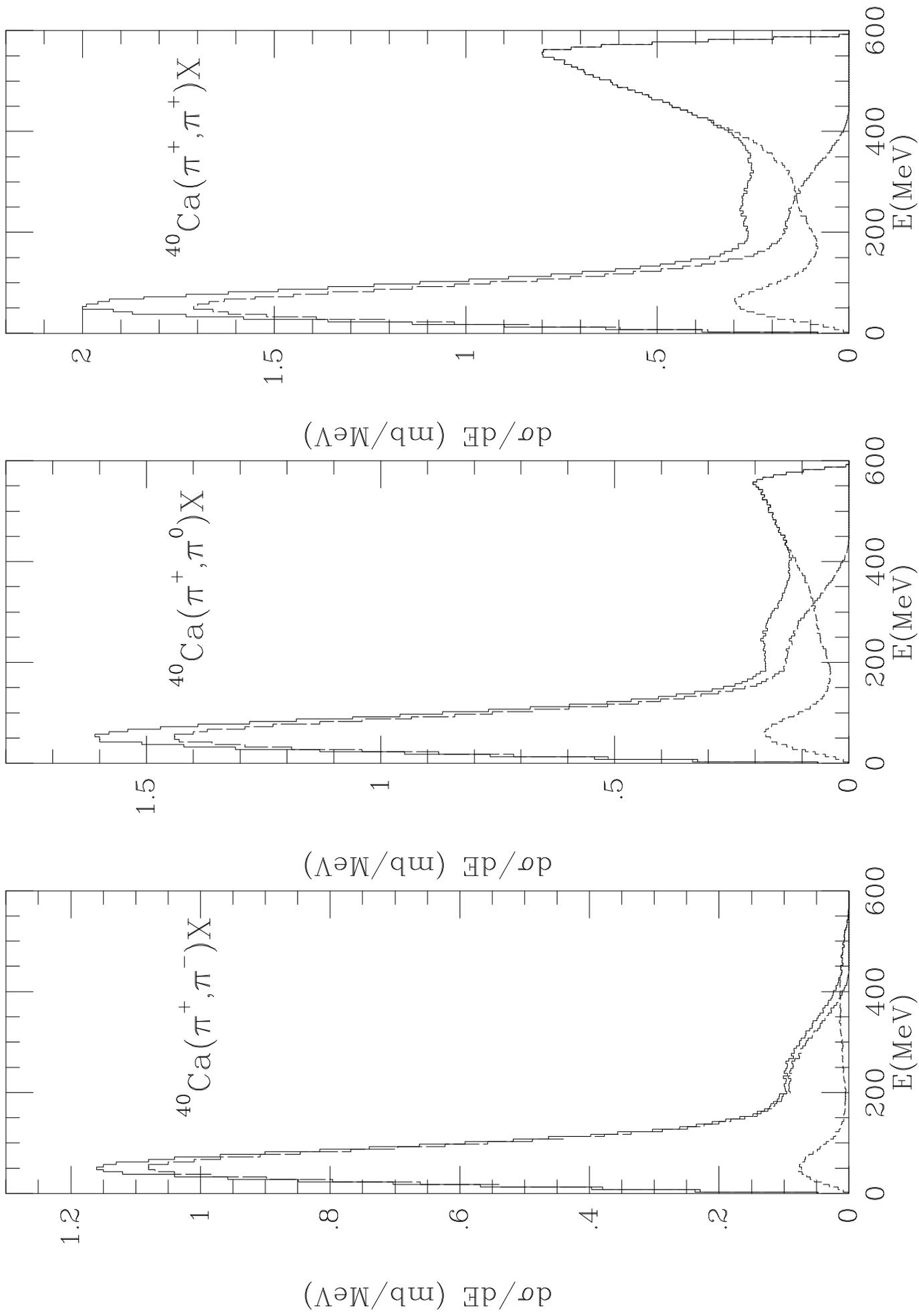
{\bf Fig. 3.}  Calculated spectra of the $^{40}Ca(\pi,\pi ')X$ reactions
at T$_{\pi}$=600 MeV. Continuous line: full model, long-dashed line:
$\pi$-production channels, short-dashed line: quasielastic channels.

\medskip
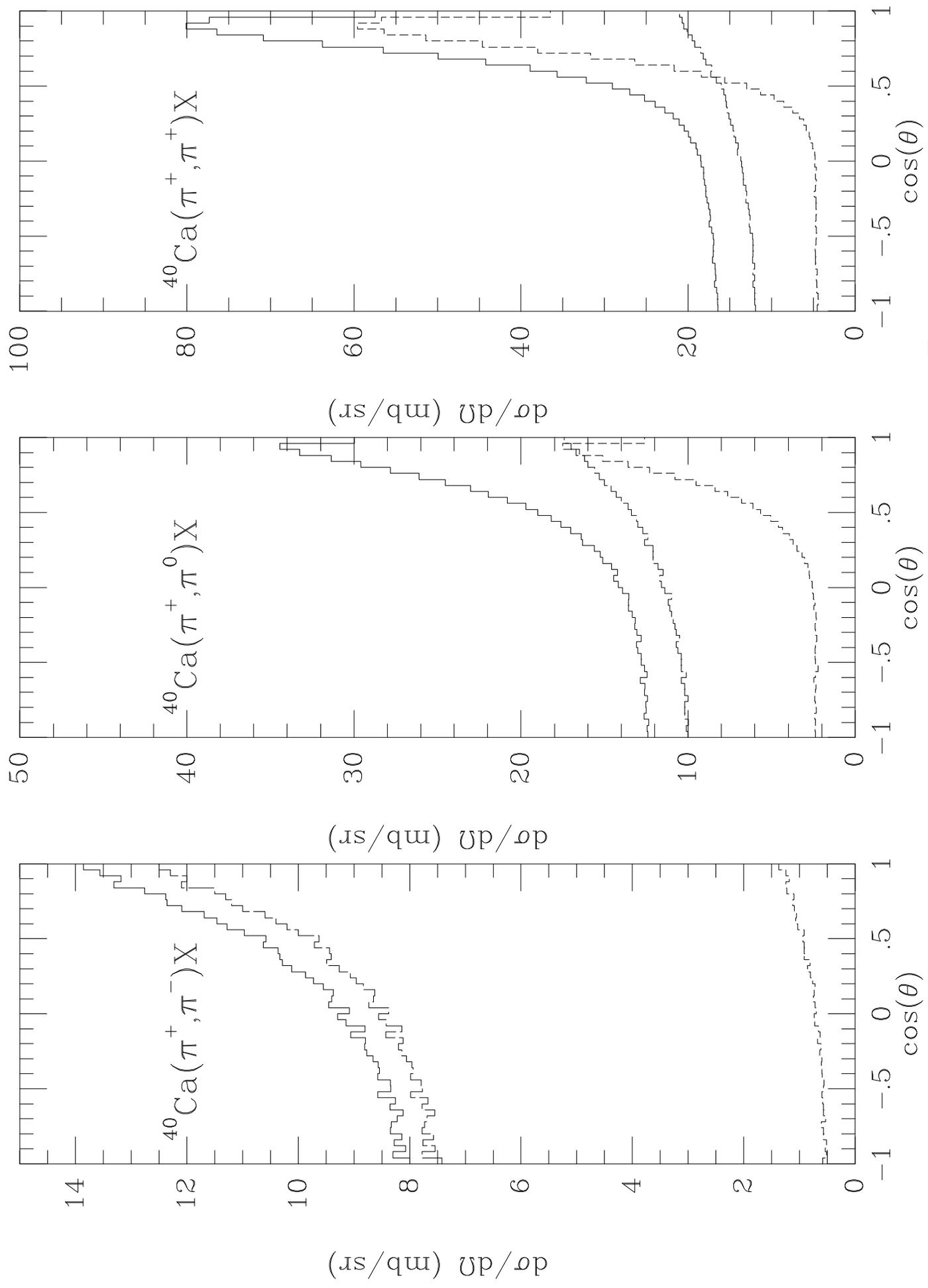
{\bf Fig. 4.}  Calculated angular distributions of the $^{40}Ca(\pi,\pi ')X$
reactions at T$_{\pi}$=600 MeV.
Continuous line: full model, long-dashed line:
$\pi$-production channels, short-dashed line: quasielastic channels.

\medskip
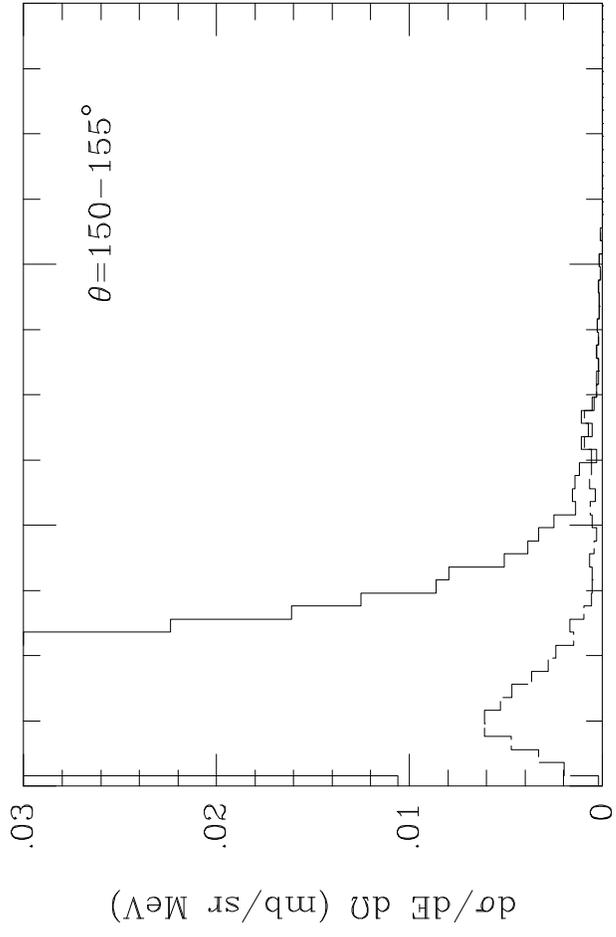
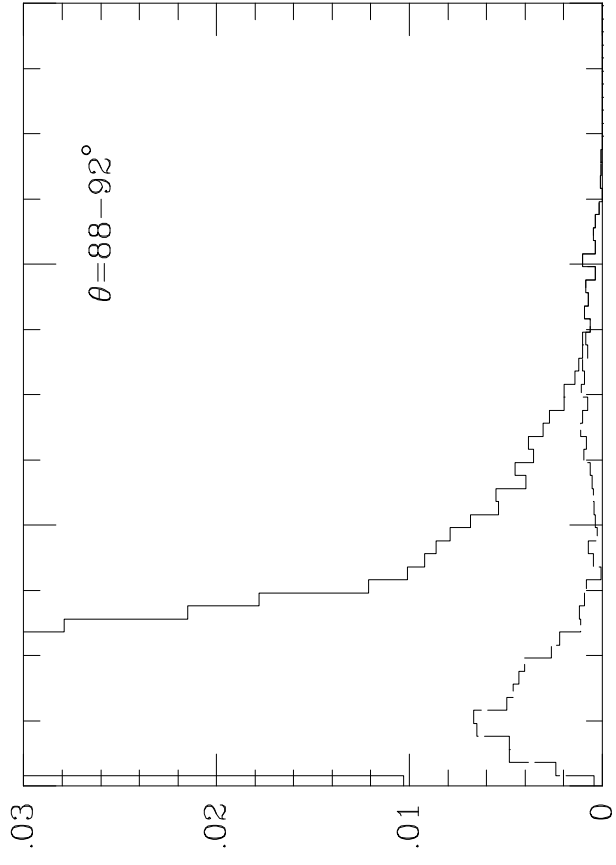
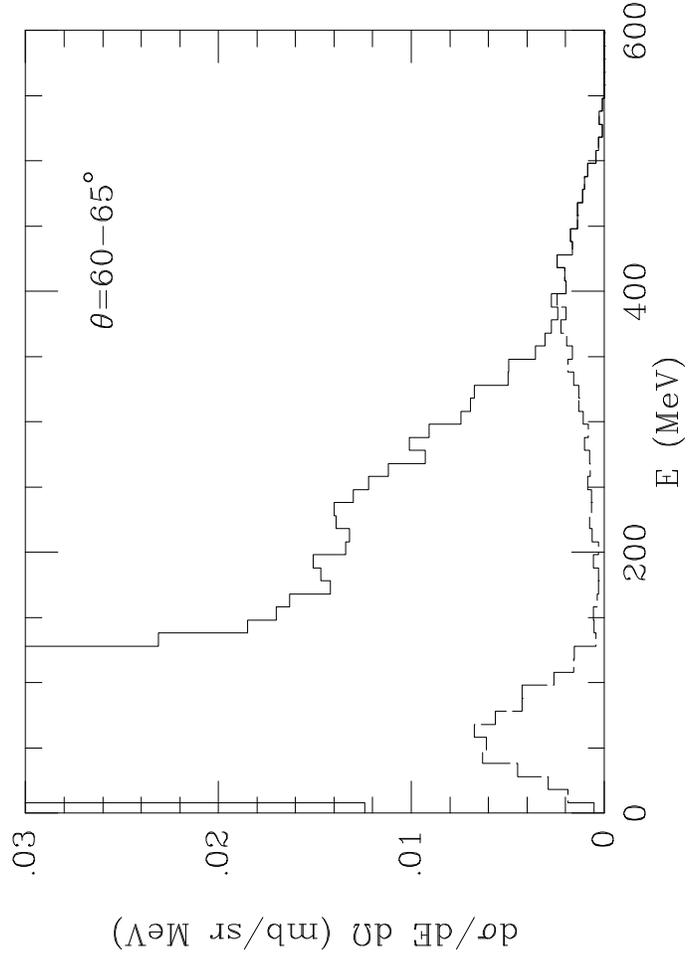
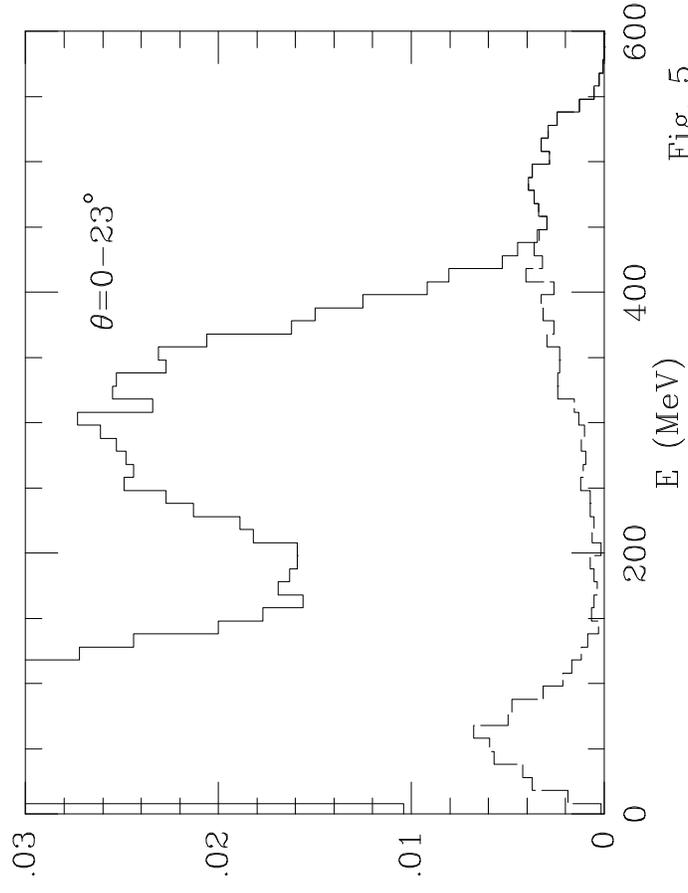
{\bf Fig. 5.}  Calculated angle-energy distributions of the
$^{40}Ca(\pi^+,\pi^-)X$ reaction
at T$_{\pi}$=600 MeV. Continuous line: full model, dashed line:
quasielastic channels.

\medskip
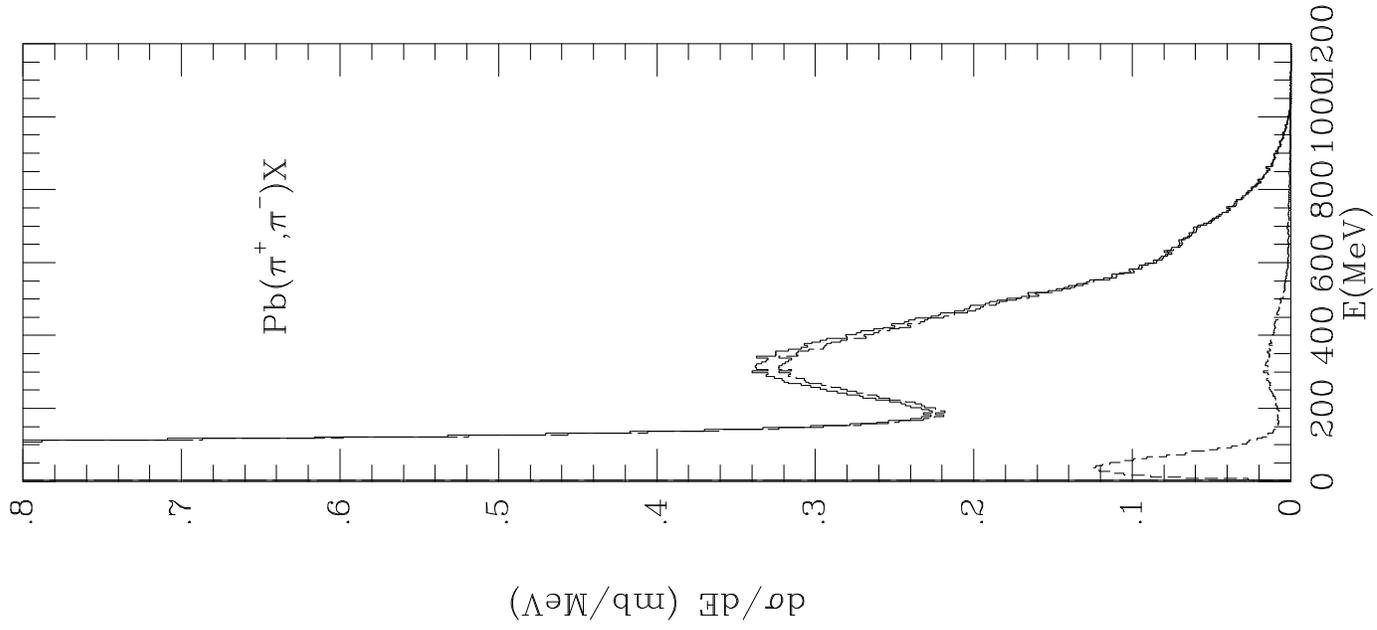
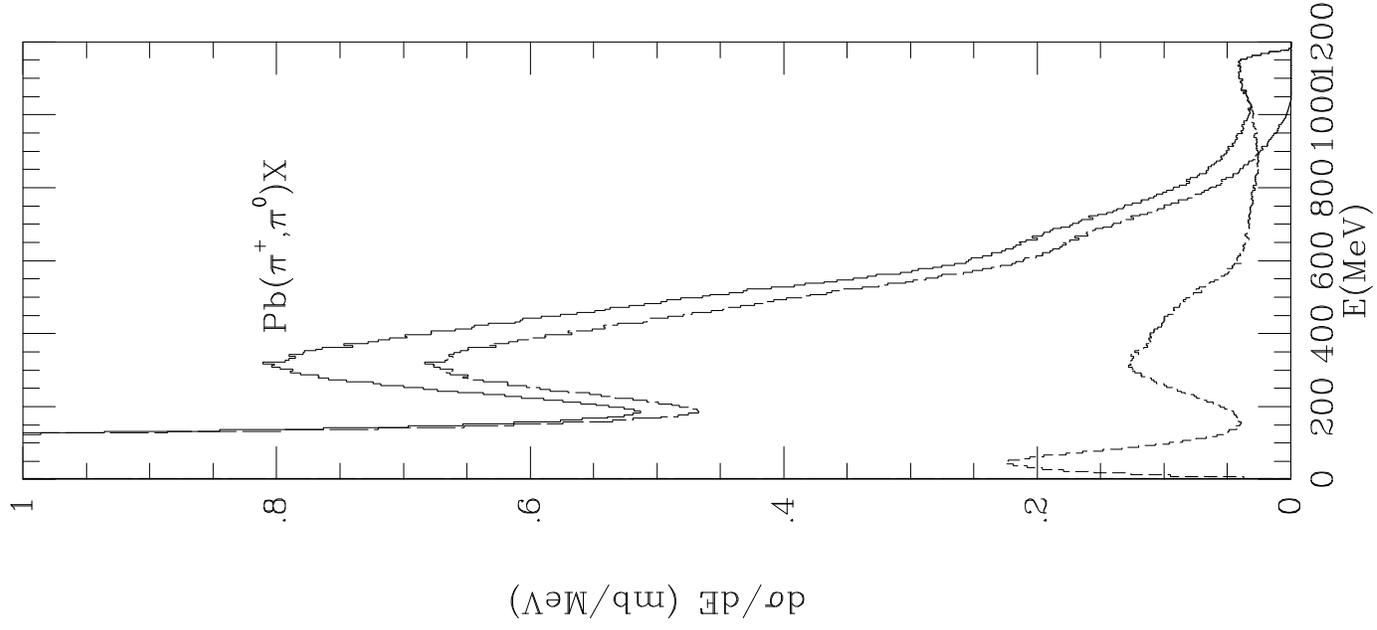
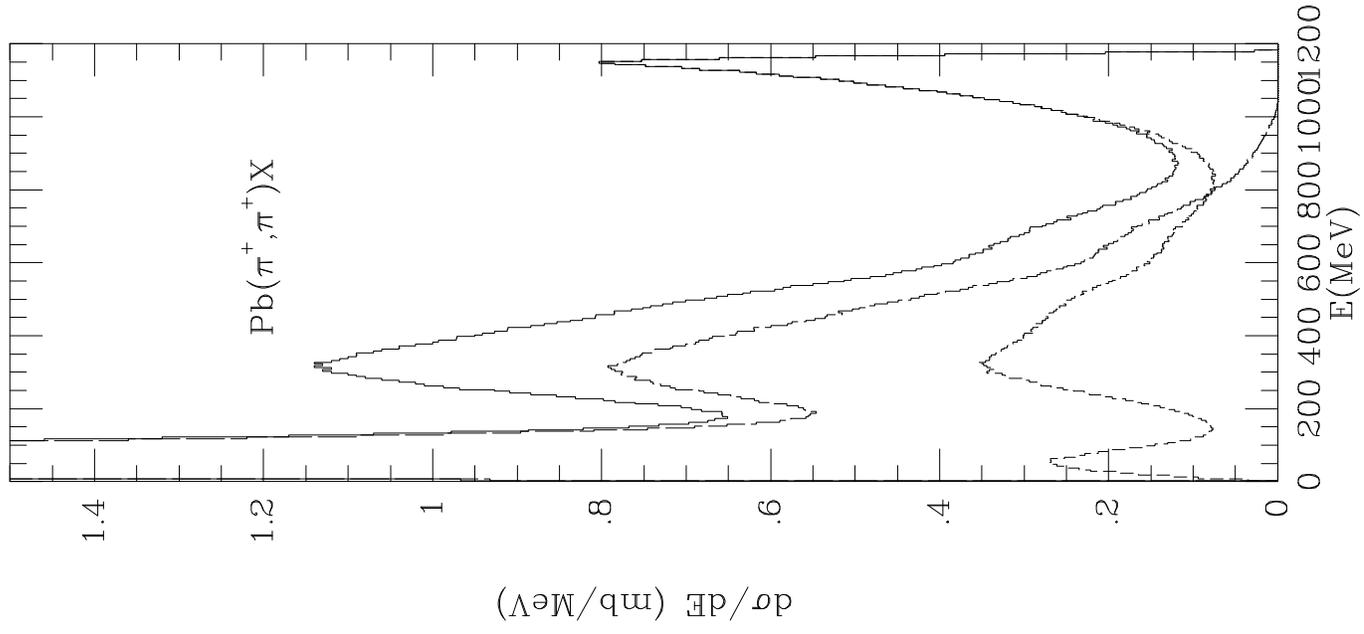
{\bf Fig. 6.}  Calculated spectra of the $^{208}Pb(\pi,\pi ')X$ reactions
at T$_{\pi}$=1200 MeV. Continuous line: full model, long-dashed line:
$\pi$-production channels, short-dashed line: quasielastic channels.

\medskip
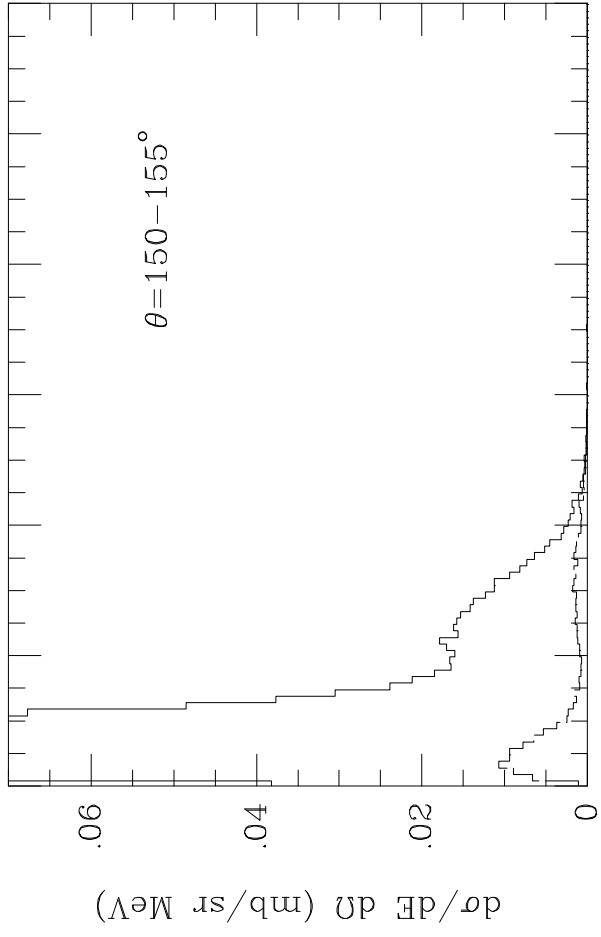
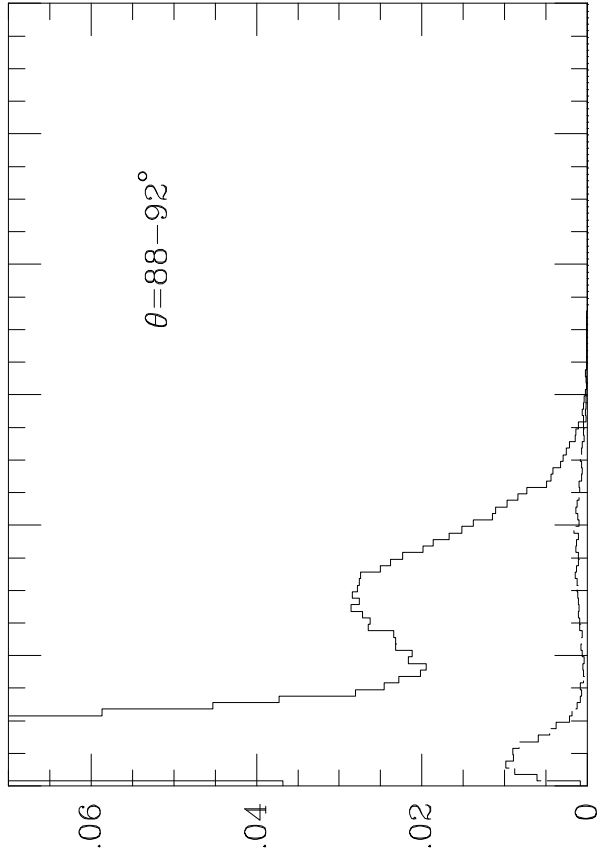
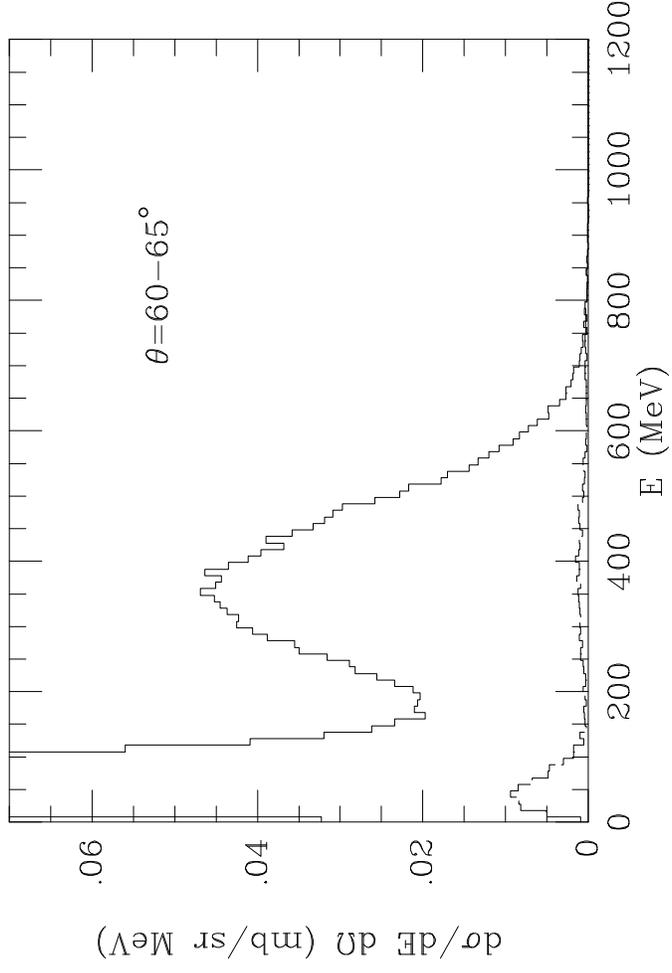
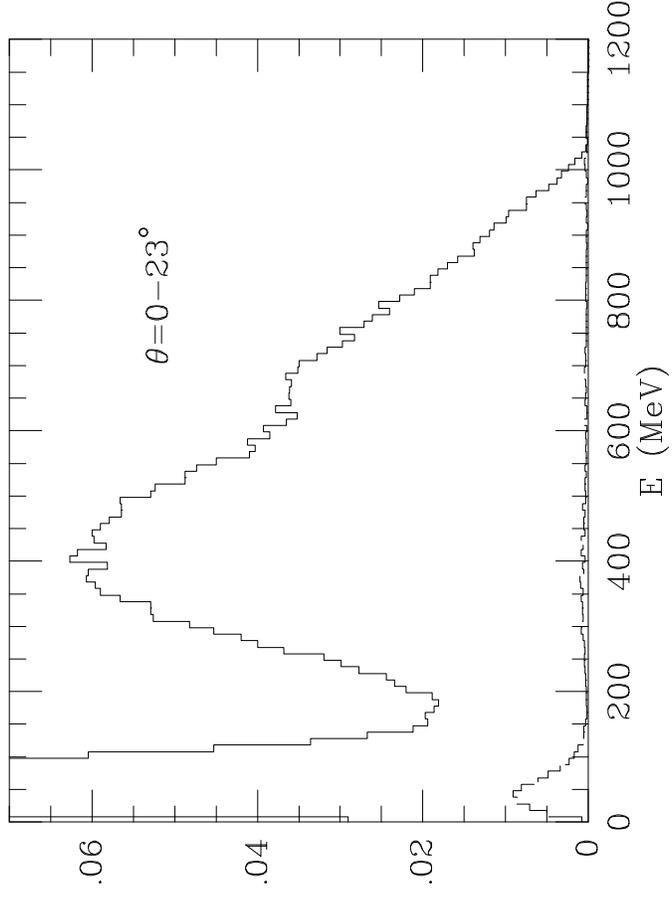
{\bf Fig. 7.}  Calculated angle-energy distributions of the
$^{208}Pb(\pi^+,\pi^-)X$ reaction
at T$_{\pi}$=1200 MeV. Continuous line: full model, dashed line:
quasielastic channels.

\medskip
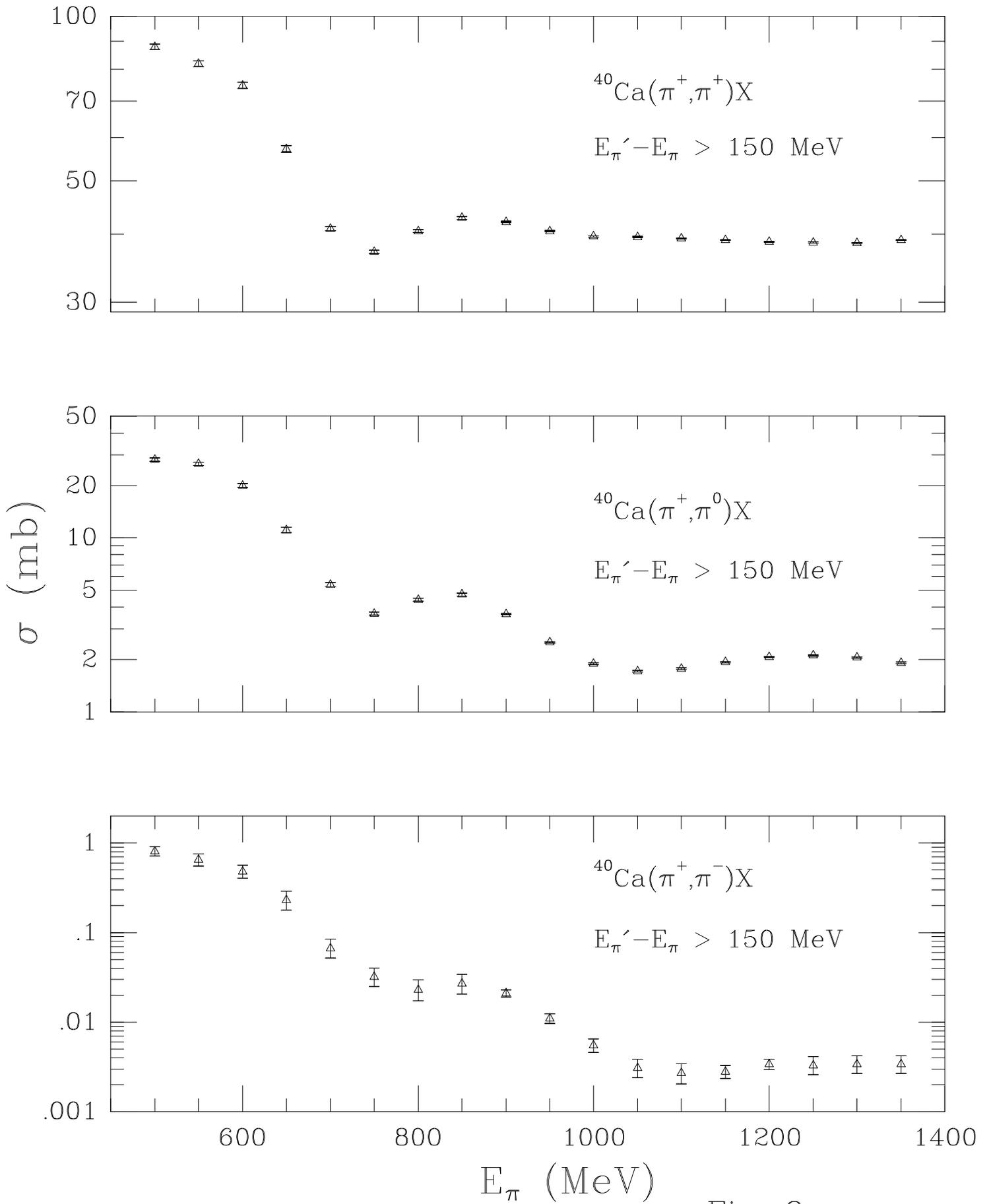
{\bf Fig. 8.}  Calculated cross section of the $^{40}Ca(\pi,\pi ')X$ reactions
as a function of the energy. Error bars represent statistical uncertainty.

\end{document}